# Local Strain Heterogeneity Influences the Optoelectronic Properties of Halide Perovskites

Timothy W. Jones[#,1], Anna Osherov[#,2], Mejd Alsari[#,3], Melany Sponseller[2], Benjamin C. Duck[1], Young-Kwang Jung[4], Charles Settens[2], Farnaz Niroui[2], Roberto Brenes[2], Camelia V. Stan[6], Yao Li[6,7], Mojtaba Abdi-Jalebi[3], Nobumichi Tamura[6], J. Emyr Macdonald[8], Manfred Burghammer[9], Richard H. Friend[3], Vladimir Bulović[2], Aron Walsh[4,5], Gregory J. Wilson[1], Samuele Lilliu[10,11], and Samuel D. Stranks[*,2,3]

[1]CSIRO Energy Centre, Mayfield West NSW 2304, Australia

[2]Research Laboratory of Electronics, Massachusetts Institute of Technology, 77 Massachusetts Avenue, Cambridge, Massachusetts 02139, USA

[3]Cavendish Laboratory, University of Cambridge, JJ Thompson Avenue, Cambridge CB3 0HE, UK

[4]Department of Materials Science and Engineering, Yonsei University, Seoul 03722, Korea

[5]Department of Materials, Imperial College London, Exhibition Road, London SW7 2AZ, UK

[6]Advanced Light Source, Lawrence Berkeley National Laboratory, Berkeley, California 94720, USA

[7]Xi'an Jiaotong University, State Key Laboratory for Mechanical Behavior of Materials, Xi'an, China

[8]School of Physics and Astronomy, Cardiff University, Cardiff CF24 3AA, UK

[9]European Synchrotron Radiation Facility, Grenoble, France

[10]Department of Physics and Astronomy, University of Sheffield, Sheffield S3 7RH, UK

[11]The UAE Centre for Crystallography, UAE

[#]These authors contributed equally

[*]sds65@cam.ac.uk

**Halide perovskites are promising semiconductors for inexpensive, high-performance optoelectronics. Despite a remarkable defect tolerance compared to conventional semiconductors, perovskite thin films still show substantial microscale heterogeneity in key properties such as luminescence efficiency and device performance[1,2,3]. This behavior has been attributed to spatial fluctuations in the population of sub-bandgap electronic states that act as trap-mediated non-radiative recombination sites.[1,4] However, the origin of the variations, trap states and extent of the defect tolerance remains a topic of debate, and a precise understanding is critical to the rational design of defect management strategies.[5,6] By combining scanning X-ray diffraction beamlines at two different synchrotrons with high-resolution transmission electron microscopy, we reveal levels of heterogeneity on the ten-micrometer scale (super-grains) and even ten-nanometer scale (sub-grain domains). We find that local strain is associated with enhanced defect concentrations, and correlations between the local structure and time-resolved photoluminescence reveal that these strain-related defects are the cause of non-radiative recombination. We reveal a direct connection between defect concentrations and non-radiative losses, as well as complex heterogeneity across multiple length scales, shedding new light on the presence and influence of structural defects in halide perovskites.**

Polycrystalline methylammonium lead iodide (MAPbI$_3$) films were solution-processed on glass cover slips or silicon substrates from lead acetate and MAI-based precursor solutions containing hypophosphorous acid additives[7] (see Methods). An SEM image of the film is shown in the inset of **Figure 1a**, revealing grain sizes of ~0.5-1 μm, along with an example of the uniquely-shaped Au particles solution-processed on the film surface to allow registering and correlating the same scan area between different experiments in this work. The films have a strong preferential orientation with the <110> and <220> planes as the primary observed reflections (see Extended Data Figure 1 for the full macroscopic X-ray diffraction (XRD) pattern). To structurally characterize the perovskite grains on the microscale we utilized the scanning micro-XRD (μXRD) beamline 12.3.2 at the Advanced Light Source (ALS) with a spatial resolution on our samples of ~2.5 μm (see Methods for further details). We held the samples under flowing nitrogen at 240 K during the measurements, which we found minimized any potential beam damage[8]. The summed 2D powder pattern for a 70×70 μm$^2$ region indexes correctly to the tetragonal phase of MAPbI$_3$ and the chi-integrated diffractogram from the entire region matches the macroscopic diffractogram (Figure 1a). The <220> and <222> are the dominant reflections in the μXRD patterns due to geometry of the experimental setup, which did not allow us to observe reflections below scattering vector $q = 1.33$ Å$^{-1}$ (Extended Data Figure 2). Line profiles show excellent match to tetragonal MAPbI$_3$ [9] and negligible traces of crystalline impurity peaks, consistent with minimal beam damage to the sample (see Extended Data Figure 3).

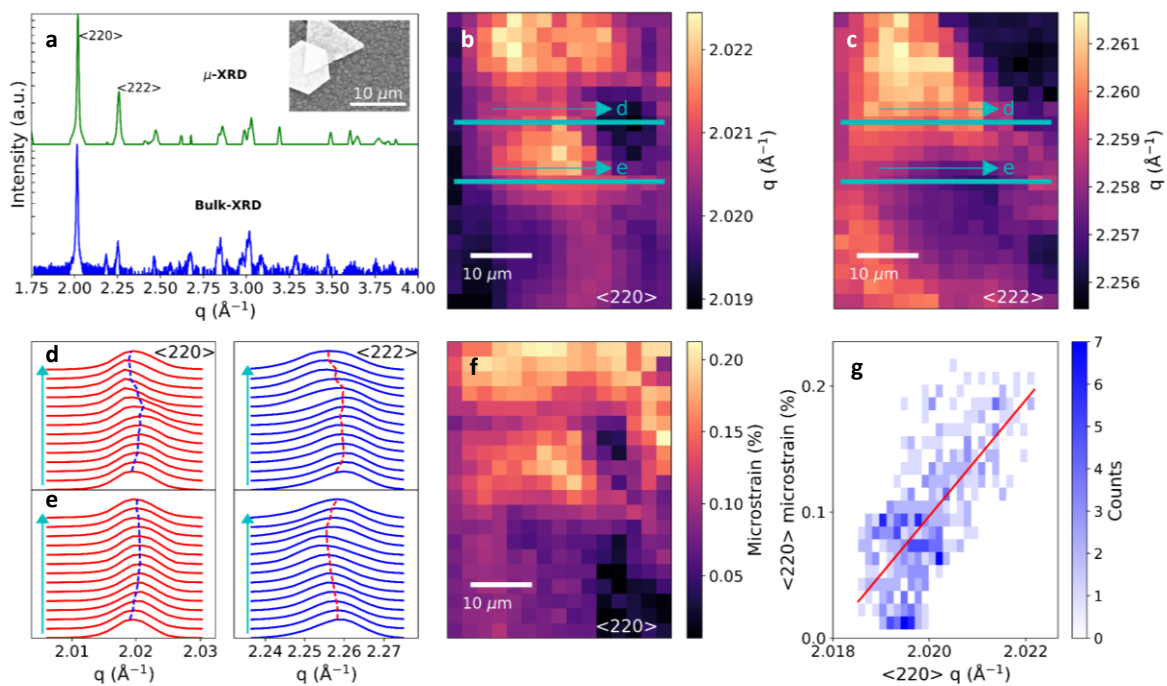

**Figure 1. Characterizing structural heterogeneity in MAPbI$_3$ films on glass cover slips by μXRD. (a)** Comparison of the macroscopic bulk XRD pattern with the micro-XRD pattern (both collected at 240 K) summed over a 70×70 μm$^2$ spatial region, with the key reflections labelled. Inset: SEM image of the perovskite grains along with ~ten-micrometer-sized Au fiducial marker particles. **(b)** Local <220> and **(c)** <222> diffraction peak $q$ maps revealing grain orientation heterogeneity. **(d)** and **(e)** Selected slices of the <220> (red) and <222> (blue) through the maps in (b) and (c) illustrating the complex strain patterns present within the film. Vertical lines indicate peak position as determined through peak profile fitting and are a guide to the eye. **(f)** Microstrain map for the <220> diffraction peak. **(g)** Histogram of the calculated microstrain and corresponding scattering vector $q$ for the <220> diffraction peak. The solid line is a linear regression fit to a scatter plot of the data, revealing a statistically-significant correlation (negligible $p$-value, see Methods).

In Figure 1b and c we show the spatial maps of the peak scattering vector $q$ of the azimuthally-integrated <222> and <220> peaks (see Extended Data Figure 4 for maps of peak intensities). These maps reveal distinct local structural heterogeneity on the scale of beam resolution, which is consistent with heterogeneous local surface grain orientations, preferred orientations or changes in level of crystallinity across the film. These distinct local patterns are evidence for regions that share similar crystallographic properties over a ten-micrometer scale (super-grains), a much larger scale than the grains observed in SEM images.

Figure 1d and e show the μXRD patterns taken from horizontal slices as indicated by the coloured lines on Figure 1b and **c**. Subtle shifts in the peak position and broadening reveal the presence of detailed microscale structural heterogeneity. This heterogeneity in the local $q$ vector for the <220> orientation of ~0.15%, corresponding to a spontaneous stress of ~19 MPa (based on a Young's Modulus of 12.8 GPa[10]), is typical of different regions of the films measured, and is a level of heterogeneity unobservable with laboratory diffraction techniques in both spatial resolution and peak variation[11]. Interestingly, the μXRD slice in Figure 1d depicts a region with long-range parallel coupling between the two reflections in local $q$ variation, whereas Figure 1e depicts a region with anti-parallel coupling between the two reflections (Extended Data Figure 5). We interpret these observations as evidence of complex local strain variations [12], and thus our results reveal long-range (>10 μm) strain patterns present throughout the polycrystalline film. Interestingly, we also find that passivating the sample through exposure to humid air and light soaking, which decreases the trap density and increases the luminescence and device performance[13], removes this coupling (Extended Data Figure 6), hinting at a relationship between local structure, defect distributions and luminescence. This observation is consistent with a recent macroscopic study proposing light-induced strain relief in films and resulting device performance enhancement[14].

After subtracting the instrumental contribution towards peak broadening of the line profiles, we consider the extreme cases of contributions from microstrain-only and crystallite-size only, as defined by the Williamson-Hall formalism[15]. The variance of the crystallite-size-only is considerably larger and thus we conclude that microstrain is the dominant contributor towards peak breadth in these samples (see Extended Data Figure 7 and Supplementary Information (SI) for further discussion). We show the resulting microstrain map of the dominant <220> reflection in Figure 1f (see Extended Data Figure 8 for <222>). This reveals that the microstrain also has a complex local heterogeneity with a typical magnitude of ~0.1-0.2 %, indicating that each grain cluster has its own local strain environment. Importantly, there is a strong correlation between $q$ and microstrain for the <220> peak (Figure 1g; Extended Data Figure 8 for the <222>). That is, XRD peaks with the larger local $q$ (lower $d$-spacing) contain the largest structural broadening due to microstrain (and vice versa). This suggests that the strain in the polycrystalline films is compressive, i.e. acting to reduce the volume of the unit cell. We observe similar correlations in an alloyed 'triple cation' $MA_{0.15}FA_{0.79}Cs_{0.06}Pb(I_{0.85}Br_{0.15})_3$ sample, suggesting this observation can be generalised to other compositions [16] (Extended Data Figure 9).

In order to further investigate the long range behaviour and its relationship to local grains, we performed scanning nanofocus XRD (nXRD) measurements at the ID13 beamline at the European Synchrotron Radiation Facility (ESRF) (Extended Data Figure 10)[9]. A MAPbI$_3$ perovskite film prepared as above was raster scanned (beam spot size 200 × 200 nm$^2$; see Methods for details). We show a quiver plot for the <110> orientation in **Figure 2a**, where the value of $\chi_p$ for each diffraction spot is represented using an arrow with its centre located in the spatial position from which the diffraction spot was acquired, and with an orientation and colour corresponding to $\chi_p$ [9]. Diffraction spots adjacent both in real and reciprocal space coordinates were considered as belonging to the same cluster, here indicated as 'super-grain' (see Methods for details). We highlight in blue and red two super-grains with the largest covered areas calculated as the number of pixels within the super-grain times the pixel area (400×400 nm$^2$). This observation of long-range features is consistent with our µXRD results.

Furthermore, the super-grains also exhibit local strain ($q$) variations within their dimensions (Figure 2b and c), which hints at complex local environments at a smaller scale than the µXRD beam footprint.

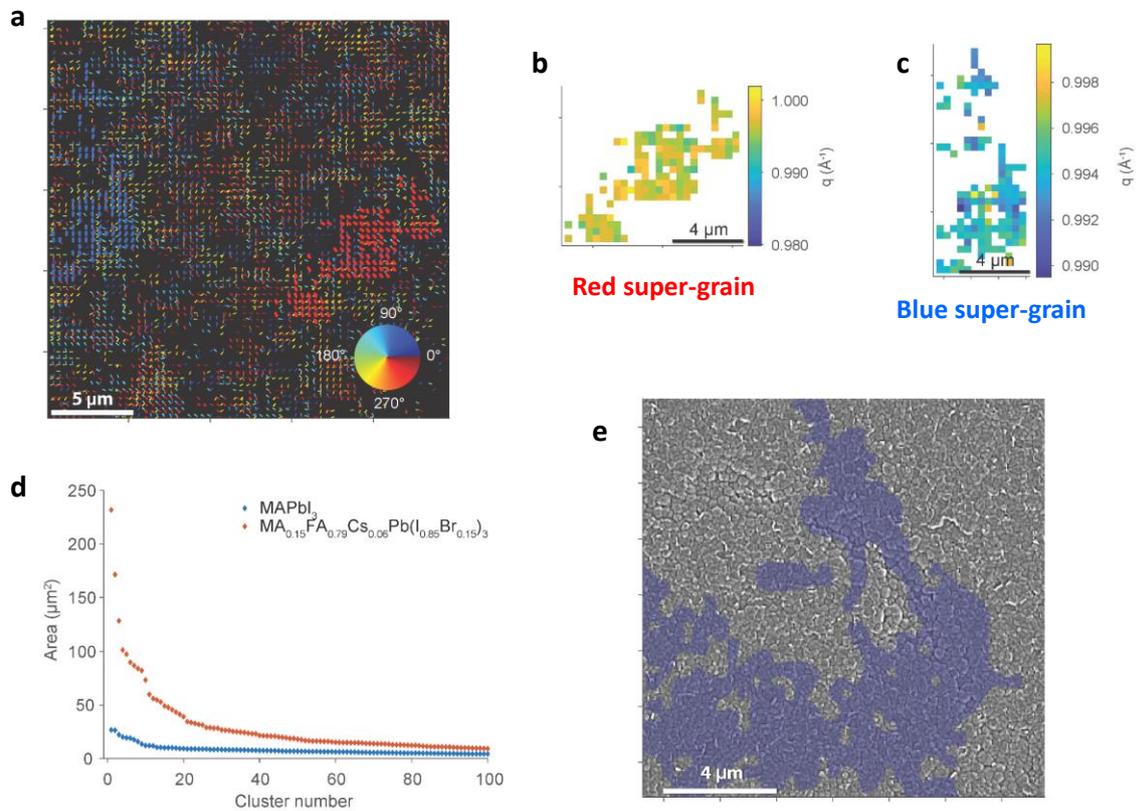

**Figure 2. Nano-XRD measurements and super-grain analysis. (a)** Quiver plot highlighting the two largest super-grains showing the <110> reflection in a MAPbI$_3$ sample deposited on a Si/SiO$_2$ substrate. **(b,c)** Local variations of the scattering vector $q$ extracted from the two super-grains in (a). **(d)** Super-grain size distributions for MAPbI$_3$ and MA$_{0.15}$FA$_{0.79}$Cs$_{0.06}$Pb(I$_{0.85}$Br$_{0.15}$)$_3$. **(e)** Overlay between the quiver plot (<210>) and SEM image in the same scan area for a MA$_{0.15}$FA$_{0.79}$Cs$_{0.06}$Pb(I$_{0.85}$Br$_{0.15}$)$_3$ sample.

The super-grain sizes for the <110> reflection are plotted in Figure 2d, showing that the largest regions cover an area of ~25 µm$^2$, extending well beyond the grain size observed in SEM images (~1 µm$^2$). We find this disparity is further exaggerated in the triple cation

MA$_{0.15}$FA$_{0.79}$Cs$_{0.06}$Pb(I$_{0.85}$Br$_{0.15}$)$_3$ samples, which show super-grains as large as ~250 μm$^2$ (Figure 2d) despite SEM grain sizes of only ~0.1 μm$^2$ ; this is clearly seen in the overlay of an SEM image and a quiver plot highlighting the largest supergrain in Figure 2e (see Extended Data Figure 11 for other supergrains). This remarkable finding sheds light onto the apparent paradox whereby 'small-grain' triple cation perovskite films still attain the highest device PCE [17]. Our results suggest that the critical grain size is actually the longer-range structural super-grains rather than the grains viewed in SEM images. The formation of these large clusters with analogous orientation and facet control will have beneficial properties for carrier transport. The results also suggest that the grain orientation is non-random, with either templated nucleation/growth or fragmentation of larger grains for instance upon drying or annealing. Such a mechanism could be akin to the long-range co-orientation of crystal plates in echinoderms [18]. These results highlight important new questions on nucleation and lateral carrier transport for the community.

To now explore crystallinity at a sub-grain resolution we turn to transmission electron microscopy (TEM). In **Figure 3a**, we show a cross-sectional bright-field TEM image of the Si|perovskite interface. The sample was prepared by thinning down the 0.4 μm long lamella that appeared as an individual grain in the top-view SEM image by focused-ion beam (FIB). A selected-area electron diffraction (SAED) pattern obtained from a 200-nm region within the lamella is outlined by a circle in the micrograph and shown in the inset of Figure 3a. The SAED pattern indicates a non-single crystalline nature of the "single grain" observed in SEM. Although the *d*-spacing corresponds to the tetragonal MAPbI$_3$ perovskite structure, the presence of elongated diffraction spots as well as a weak diffraction ring is a strong indicator of imperfections within the lattice that likely originate from strain and/or other defects.

To probe crystallinity at a deeper scale, a high-resolution TEM (HR-TEM) micrograph was collected from a 70×70 nm$^2$ region (Figure 3b). The micrograph shows a lack of lattice continuality in the tested area as indicated by the presence of domains and structural defects. Fast Fourier Transform (FFT) patterns generated from various 10×10 nm$^2$ regions of the HR-TEM micrograph are outlined by the coloured boxes in Figure 3c and clearly demonstrate structural heterogeneity *within* a single grain. The regions marked by black and purple boxes possess near-identical diffraction patterns and are highly crystalline as indicated by the sharp diffraction spots. The identical *d*-spacing of the diffraction spots indicate a similarity in crystallinity and sub-grain crystallite orientation on the 10×10 nm$^2$ scale. The highly crystalline regions marked by the black and purple boxes contrasts with regions bound by the red and blue boxes. The markedly different patterns indicate the whole grain is not uniformly crystallized. The blue region is well-crystallized but shows the presence of more than one diffraction pattern. This might be a result of multi-grain overlap as well as indicate the presence of various structural defects such as a micro-twins or/and dislocations [19]. In contrast, the red box shows a much weaker pattern indicating a poorly-crystallised or amorphous region within the same grain. While amorphization due to beam damage cannot be ruled out completely, a lack of homogeneity in the amorphization signature through the sample makes irradiation-induced amorphization less plausible[20].

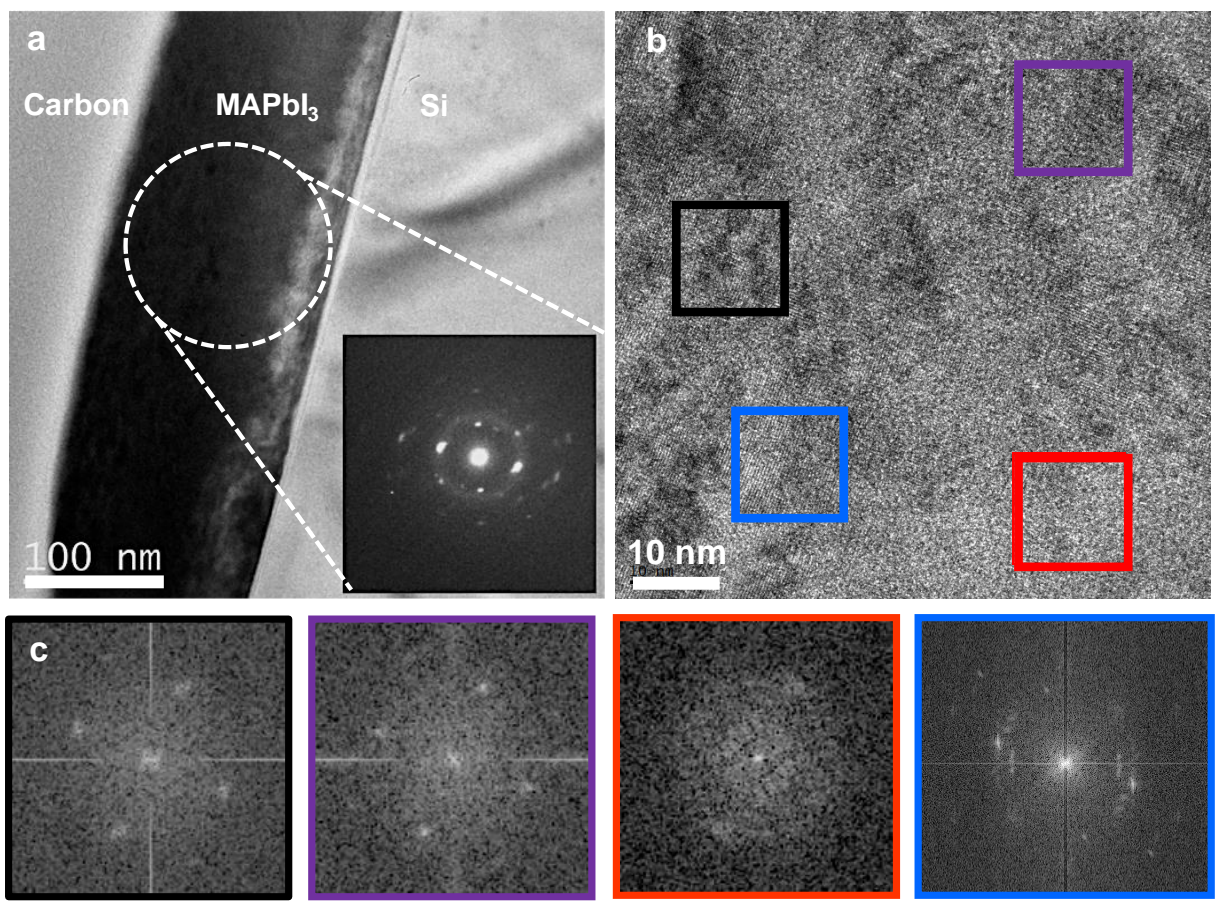

**Figure 3. TEM electron diffraction analysis of MAPbI₃ perovskite grains**. **(a)** Cross-section image of single perovskite grain on Si <100> substrate and selected area electron diffraction (SAED) pattern collected from the region indicated (inset). **(b)** Higher-magnification TEM and **(c)** Fast Fourier transform (FFT) patterns generated across different regions of the grain shown in (b).

These results suggest that we need to re-think the conception of a perovskite grain as a single crystalline entity, for example as viewed in a SEM. Each of these entities are in fact comprised of many sub-crystallites on a ~10-100 nm scale above that of the unit cell but below that of a single grain. This scale of heterogeneity is consistent with recent reports showing lower symmetry domains below 20 nm [21] and substantial spatial variation in the photo-response of polycrystalline perovskite devices even within each grain [3, 22, 23]. Further increasing complexity of the microstructure, our combined diffraction analysis demonstrates the existence of local grain clusters (super-grains) with shared crystal orientation and heterogeneous strain on the scale of tens of micrometers. We propose that this level of structure is imparted on the film during the stage of nucleation, where differences in local substrate morphology or local concentration differences during solution deposition template the growth of certain crystal planes.

We now explore the impact of a strained crystal on local point defect concentrations. Using a first-principles atomic model, we introduce compressive <110> strain with magnitude on the order observed in our local measurements (~0.2%) and probe the effect on the defect thermodynamics (**Figure 4d**). With compressive strain, there is an increase in the charged iodide vacancy ($V_I^+$) concentration by a factor of 2 with respect to the unstrained crystal due to a negative defect pressure (see Methods and Extended Data Figures 12 and 13). A spatial map of the relative defect density (i.e. ratio relative to an unstrained grain) is shown in Figure 4a. Here, we determined the local compressive strain from µXRD using the relative shift of the peak $q$-value at each local point from the minimum $q$ in the distribution (i.e. strain = ($q_{min}-q$)/$q_{min}$). Compressively strained grains are strongly associated with increased concentrations of charged vacancies compared to unstrained grains, where the soft nature of the perovskite lattice means the structure may deform readily in the presence of a defect. Indeed, the observation of a light-induced relief of local strain patterns by passivation of defect sites (cf. Extended Data Figure 6) indicates that the local lattice contraction is caused by a defect (e.g. vacancy), rather than the other way around.

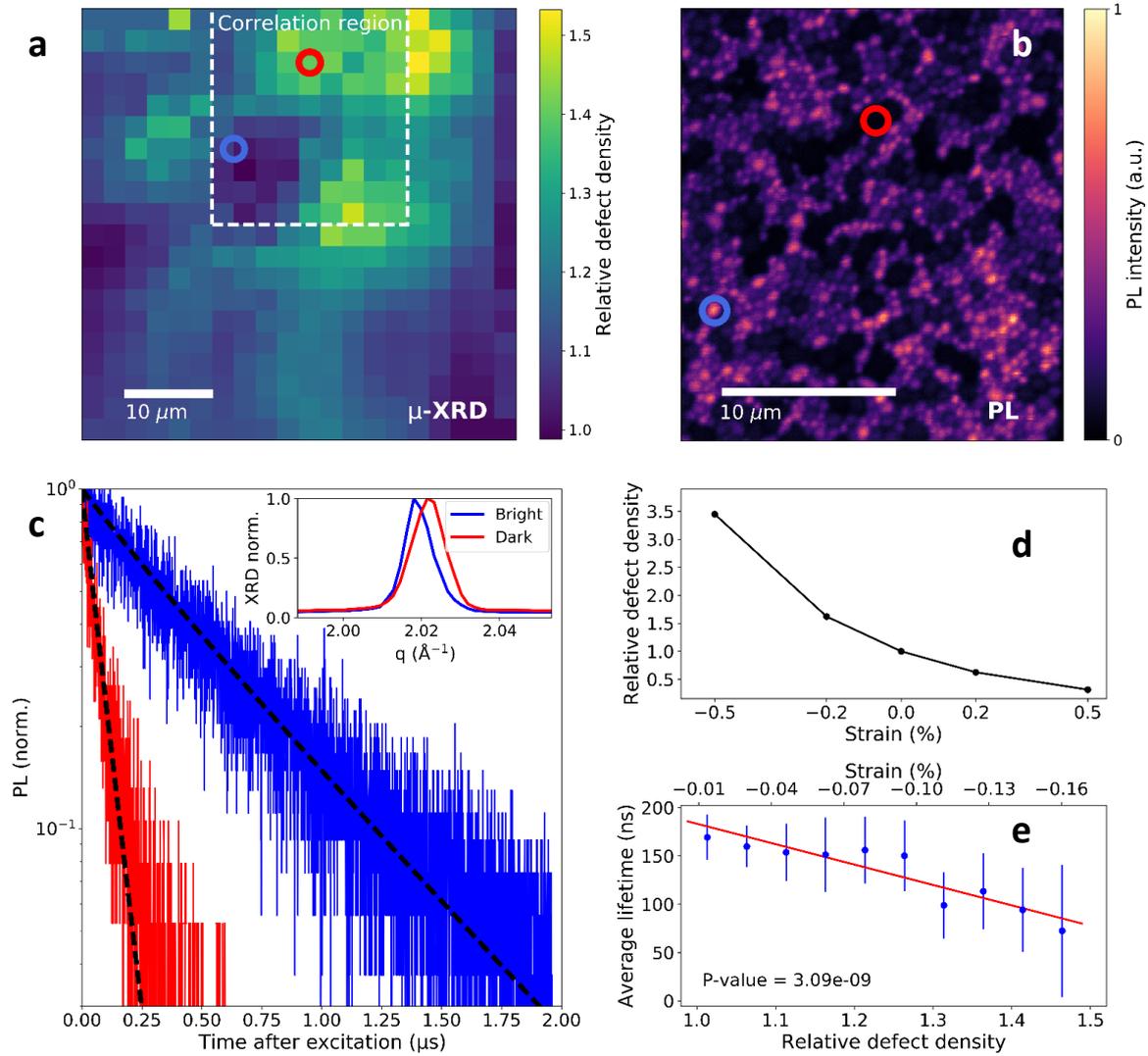

**Figure 4. Correlating the local structural and time-resolved luminescence properties of MAPbI$_3$ films.** **(a)** Spatial map of the ratio in concentration of charged iodide vacancies ($V_I^+$ defects) in <110> strained perovskite crystals to an unstrained crystal (calculated relationship shown in panel d). The dashed line denotes the correlation region between μXRD and PL. **(b)** Confocal PL intensity map of a MAPbI$_3$ perovskite film with pulsed 405-nm excitation (0.5 MHz repetition rate, 0.1 μJ/cm$^2$/pulse). **(c)** Time-resolved PL decays of the bright (blue) and dark (red) regions highlighted in (b). The solid lines are fits to the data using a trap model to extract the electronic trap density[24]. Inset: Highlighted <220> peak diffraction pattern for the bright and dark PL regions. **(e)** Scatter plots of statistically-significant

correlations between local PL lifetime and relative defect density (compressive strain). See Extended Data Figure 18 for example PL decays.

Structural defects have a strong influence on the charge-carrier lifetime and recombination in semiconductors and charged point defects typically make efficient non-radiative recombination centers due to the long-range Coulomb attraction[25]. To directly assess the impact of the observed strain-related defects on the perovskite optoelectronic properties, we correlate confocal PL measurements with μXRD measurements on the same scan area. To directly compare the two measurements, we use the Au fiducial markers and an image analysis algorithm for image registration, while also accounting for the spatial resolution differences between the measurements (see SI and Extended Data Figs 14-17 for details). We show in Figure 4b a confocal PL intensity map of the correlation region highlighted in the defect density map in Figure 4a. We note that here we are at sufficiently low excitation fluence that the local PL distribution is dominated by trap states rather than diffusion of carriers out of the local region [26]. One striking observation is the presence of regional "clustering" of bright and dim grains. That is, grains are more likely to share similar fractions of radiative and non-radiative recombination pathways with their nearest neighbours. Local clustering of defects in strained regions is consistent with the μXRD results above, again suggesting local structuring (super-grains) on a scale above that of the individual grain. We show the local time-resolved PL measurements of a bright cluster and a dim cluster in Figure 4c representing the recombination of charge carriers. The bright regions have a longer PL lifetime than the dark regions, which is consistent with increased fractions of trap-limited recombination in the latter. We extract trap densities representative of the regions by fitting the decays with a kinetic model developed previously [24], quantifying the reduction in trap density from the dark ($7.5 \times 10^{16}$ cm$^{-3}$) to the bright ($1 \times 10^{16}$ cm$^{-3}$) regions.

We show in the inset of Figure 4b the <220> μXRD peaks corresponding to the region with bright emission (long PL lifetime) and the region with dark emission (short PL lifetime). We find that the region with inferior emission intensity and carrier lifetime dynamics corresponds to a region with compressive-strained <220> lattice planes (i.e. larger $q$ and increased peak broadening), whereas the region showing brighter emission and longer carrier lifetime is comparatively unstrained (i.e. smaller $q$ and peak widths close to the instrumental broadening). We see these trends appear consistently across the correlated maps; scatter plots of the relevant quantities across the map after accounting for differences in resolution between the techniques reveal a statistically-significant decrease in PL lifetime with <220> compressive strain and, in turn, with the calculated defect concentration ratio (Figure 4e) (see Extended Data Figure 19 for microstrain).

These results have profound ramifications on our understanding of the impact of crystallinity on carrier recombination: strained grains are associated with increased fractions of charged defects and increased non-radiative decay. The relative increase in trap density for a dark grain ascertained from the trap model (Figure 4c) is consistent with the calculated magnitude of the increase in strain-related halide vacancy concentrations after considering contributions from each direction (Figure 4d). These microscale observations provide insights into the processes underpinning recent macroscopic measurements suggesting that reduced strain leads to enhanced macroscopic PL, stability and charge carrier properties [12, 14, 27, 28]. Our results show for the first time that the observation of local PL heterogeneity is substantially influenced by locally-heterogeneous strain distributions that are associated with halide vacancies.

Our work has revealed that halide perovskites feature multiple length scales of structural heterogeneity throughout the entire film, from long-range super-grain clusters, to grain-to-grain and sub-grain nanoscale variations. Their exceptional performance[29] in spite of so many layers of disorder is remarkable, and is seemingly in contrast to established semiconductor physics. Indeed, the behaviour is akin to liquid metals, which have disordered structures, yet maintain excellent charge transport properties [30]. These observations also explain the high open-circuit voltages in devices even in the earliest stages of disordered crystallite formation[31]. An interesting question arises about whether their exceptional performance is in fact *as a result of* the length scales and topology of the disorder. This needs to be squared with the unwanted presence of strain-related defects that introduce microscale non-radiative losses. Future work will be required to consolidate our understanding of the complex relationships between physicochemical and optoelectronic properties, which would ultimately guide growth of films with optimal structural properties across all scales.

**Methods**

*Sample Preparation*

All organic-based precursor salts were purchased from Dyesol. Unless otherwise stated, all other materials were purchased from Sigma-Aldrich. All film samples were prepared in and then stored in a nitrogen-filled glovebox until used. Prior to measurement, the samples were stored in a dry air desiccator for 24 hours to stabilize any photo-brightening effects.

Glass cover slip (PL and micro-XRD measurements) or Si<100> substrates (nano-XRD, TEM measurements) were washed sequentially with soap (Micro 90), de-ionized water, acetone, and isopropanol, and finally treated under oxygen plasma for 10 minutes. Thin films of $MAPbI_3$ were solution-processed by employing a methylammonium iodide (MAI) and lead acetate $Pb(Ac)_2 \cdot 3H_2O$ precursor mixture with a hypophosporous acid (HPA) additive [7]. MAI and $Pb(Ac)_2 \cdot 3H_2O$ were dissolved in anhydrous N,N-dimethylformamide at a 3:1 molar ratio with final concentration of 37 wt% and HPA added to a HPA:Pb molar ratio of ~11%. The precursor solution was spin-coated at 2000 rpm for 45 seconds in a nitrogen-filled glovebox, and the substrates were then dried at room temperature for 10 minutes before annealing at 100°C for 5 minutes.

The triple-cation-based perovskite $MA_{0.15}FA_{0.79}Cs_{0.06}Pb(I_{0.85}Br_{0.15})_3$ was prepared by dissolving $PbI_2$ (1.2 M, TCI), FAI (1.11 M), MABr (0.21 M) and $PbBr_2$ (0.21 M, TCI) in a mixture of anhydrous DMF:DMSO (4:1, volume ratios) followed by addition of 5 volume percent from CsI (TCI, Japan) stock solution (1.5 M in DMSO). We then spin-coated the perovskite solution using a two-step program at 2000 and 4000 rpm for 10 and 40 seconds, respectively, and dripping 150 µL of chlorobenzene after 30 seconds. We then annealed the films at 100°C for 1 hour.

*Fiducial Markers*

The Au nanoplates were synthesized following a modified procedure reported by Gu et al. [32]. A 25 mL solution of ethylene glycol with 0.054 mmol of $HAuCl_4 \cdot 4H_2O$ was heated to 65°C in a water bath for 20 min. Then, 0.1 M aniline solution in ethylene glycol was added to the heated $HAuCl_4$ solution under mild stirring to acquire a 2:1 molar ratio of aniline to Au. This reaction was allowed to proceed for 3 hours without stirring leading to formation of triangular and hexagonal Au nanoplates with spherical nanoparticles as a byproduct. The nanoplates precipitated at the bottom of the vial while the supernatant mainly contained the spherical particles. The supernatant was gently removed without disturbing the precipitant and replaced with fresh ethanol. This was followed by two more rounds of precipitation to thoroughly clean the nanoplates in ethanol. Lastly, the ethanol was removed and the particles were dispersed in chlorobenzene (~0.5-1 mL). The solution was sonicated for 30 seconds and spincoated on top of the perovskite samples at 1000 rpm for 60 seconds in atmospheric conditions, resulting in a dispersion on the surfaces with a concentration of ~1 particle cluster per 100 x 100 $\mu m^2$.

*Bulk X-Ray Diffraction Measurements*

Bulk X-Ray Diffractograms (cf. Extended Data Figure 1) were collected from thin film samples using a PANalytical X'Pert Pro Multi-Purpose Diffractometer operated at 45 kV and 40 mA (Cu K-alpha radiation- 1.5418 Å) in Bragg-Brentango geometry. An Oxford Cryosystems PheniX cryostat was employed to measure while holding the sample at 240 K.

*Micro(Confocal)-Photoluminescence Measurements*

Confocal photoluminescence (PL) maps were acquired using a custom-built time-correlated single photon counting (TCSPC) confocal microscope (Nikon Eclipse Ti-E) setup with a 100X oil objective (Nikon CFI PlanApo Lambda, 1.45 NA). The cover slip samples were photo-excited through the glass-side using a 405 nm laser head (LDH-P-C-405, PicoQuant GmbH) with pulse duration of <90 ps, fluence of ~0.1 µJ/cm$^2$/pulse, and a repetition rate of 0.5 MHz. The PL from the sample was collected by the same objective and the resulting collimated beam passes through a long-pass filter with a cut-off at 416 nm (Semrock Inc., BLP01-405R-25) to remove any residual scattered or reflected excitation light. A single photon detecting avalanche photodiode (APD) (MPD PDM Series 50 mm) was used for the detection. The sample was scanned using a piezoelectric scanning stage. The measurements were acquired and intensity and average-lifetime maps extracted using the commercial software SymphoTime 64 (PicoQuant GmbH).

*Micro-X-Ray Diffraction Measurements and Analysis*

X-Ray microdiffraction data at the Advanced Light Source were collected on beamline 12.3.2 [33]. The samples were placed in reflective geometry at a 15° grazing incidence angle (Extended

Data Figure 2); we account for the associated increased beam footprint (2.5 μm × 9.6 μm) on the sample in our correlations. The substrate was cooled to 240 K as measured by a thermocouple with a Cryostream 700 Series from Oxford Cryosystems. The sample was protected from condensation by a protective sheath of dry nitrogen evaporate. The 2D X-ray diffraction patterns were collected using a DECTRIS Pilatus 1 M hybrid pixel area detector placed at an angle of 50° with respect to the incident beam and at a distance about 160 mm from the sample. The energy of the incident beam was set to 10 keV (wavelength λ = 1.2398 A). The diffraction patterns were analyzed using the X-Ray Microdiffraction Analysis Software (XMAS) package[34]. Experimental geometry was calibrated using an $Al_2O_3$ powder, in which the detector position and angular tilts with respect to the incident x-ray beam and sample were refined. An additional calibration on the sample themselves were performed on some of the data to minimize displacement errors due to the shallow incident angle of the x-ray beam onto the sample surface. The 2D diffraction patterns were then integrated over the azimuthal direction to obtain 1D diffractograms, from which radial peak profiles were obtained and strain and particle size were calculated. See Extended Data Text for details of further data extraction, fitting, image registration and correlation fits.

*Nano-X-Ray Diffraction Measurements and Analysis*

Scanning nanofocus XRD measurements were conducted at the ID13 beamline at the European Synchrotron Radiation Facility. Samples were mounted on an xyz piezo stage and were illuminated in transmission geometry (Extended Data Figure 10) with a monochromatic beam (λ ≈ 0.8377 nm, spot size ~0.2 × 0.2 μm$^2$). Diffraction images were collected using an Eiger X 4M (Dectris AG, Switzerland) detector with 2168 row pixels and 2070 column pixels (75 × 75 μm$_2$ pixel size) and an exposure of 0.1 s per diffraction pattern (acquisition time of 0.1 s per diffraction pattern). The detector was placed 19.41 cm away from the sample. The detector position and geometry were calibrated by recording a diffraction pattern of the standard calibration material corundum ($\alpha$-$Al_2O_3$). Samples were raster scanned over a $40 \times 40$ μm$^2$ area with a step size of ~400 nm. Measurements were performed under ambient conditions (lab temperature ~24°C, relative humidity ~40%). The collected data consisted of 10201 diffraction images per scan area.

Diffraction data was analysed as in ref [9]. Briefly, after the construction of an average diffraction pattern for the diffraction collected across the entire scan area for the $MAPbI_3$ and $FA_{0.83}Cs_{0.17}Pb(I_{0.8}Br_{0.2})_3$ films, we create circular regions of interest (ROIs) around the <110> and <210> perovskite reflection rings, respectively. Diffraction spots in the ROI-restricted pattern of the scan were analyzed and used to perform grain clustering based on the azimuthal angle coordinate $\chi_P$ extracted from the center of each diffraction spot. We assume that diffraction spots that are adjacent both in spatial coordinates and in reciprocal space coordinates originate from the same grain. This clustering is performed based on the pairwise Euclidean distance between pixels. We used an empirically-determined threshold value for cutting the hierarchical tree. The dataset for the identification of the diffraction spots is bi-dimensional (pixel X and pixel Y), and the dataset for the identification of the super-grains is tri-dimensional (2 spatial coordinates X and Y, and the azimuthal angle). The area of each 'super-grain' is simply calculated as the number of pixels within the 'super-grain' multiplied by the area of a pixel in microns. This gives a collection of super-grains (clusters) that can be ordered in decreasing size, which is plotted in Figure 2b. Spatial maps of the q position were obtained from the line profiles extracted for all the diffraction spots in the ROI for each super-grain.

*Transmission Electron Microscopy*

TEM and HRTEM were carried out using a JEOL FasTem-2010 instrument operating at 200 kV. Film thickness was measured from TEM cross sections. Selected area electron diffraction (SAED) patterns were obtained by positioning the selected area diffraction aperture of the TEM over the desired area of the film and/or substrate.

*First-principles simulations*

Calculations were performed on MAPbI$_3$ in the room-temperature tetragonal phase. First the unit cell (lattice vectors and internal positions) was optimized to within a force tolerance of 0.001 eV/Angstrom as calculated using density functional theory (PBEsol functional including scalar-relativistic effects) within the code VASP. The valence electrons were expanded in a plane-wave basis set with a cutoff of 700 eV, and the k-point sampling was set to 6x6x4.

The equilibrium crystal structure was then subject to compressive and tensile strain (up to 0.5%) along the <110> and <111> crystallographic orientations. Next, the formation of iodine vacancy defects was probed in a series of 5 structures for each value of strain ($\epsilon$), with the defect energy calculated from:

$$\Delta E_{D,\epsilon} = E[\text{defect}]_\epsilon - E[\text{bulk}]_\epsilon$$

The change in defect concentrations in strained and unstrained regions at finite temperatures were calculated from:

$$\frac{n_{D,\epsilon}}{n_D} = \frac{e^{-\frac{\Delta E_{D,\epsilon}}{k_B T}}}{e^{-\frac{\Delta E_D}{k_B T}}} = e^{-\frac{\Delta E_{D,\epsilon} - \Delta E_D}{k_B T}}$$

**Acknowledgments:**

TWJ would like to acknowledge the Australian Renewable Energy Agency for receipt of a post-doctoral Fellowship. GJW would like to acknowledge the support of a CSIRO Julius Career Fellowship. TWJ and GJW are grateful to the International Synchrotron Access Program for travel support. This project has received funding from the European Union's Seventh Framework Programme (FP7/2007-2013) under REA grant agreement number PIOF-GA-2013-622630, the European Research Council (ERC) under the European Union's Horizon 2020 research and innovation programme (HYPERION, grant agreement No 756962), and the Royal Society and Tata Group (UF150033). This work made use of the Shared Experimental Facilities supported in part by the MRSEC Program of the National Science Foundation under award number MDR – 1419807. This work was supported in part by the Yonsei University Future-leading Research Initiative of 2017-22-0088. This research used resources of the Advanced Light Source, which is a DOE Office of Science User Facility under contract no. DE-AC02-05CH11231. We thank the European Synchrotron Radiation Facility (ESRF) for awarding beamtime at the ID13 beamline and the ID13 beamline staff for support with the measurements. M.A.J. thanks Nava Technology Limited and Nyak Technology Limited for their funding and technical support. AO would like to acknowledge the support from the NSF under Grant No. 1605406 (EP/L000202). M.A. received funding from 'The President of the UAE's Distinguished Student Scholarship Program (DSS), granted by the Ministry of Presidential Affairs'. The authors thank David Cahen and Gary Hodes for fruitful discussion.
**References**

1. deQuilettes DW, Vorpahl SM, Stranks SD, Nagaoka H, Eperon GE, Ziffer ME, *et al.* Impact of microstructure on local carrier lifetime in perovskite solar cells. *Science* 2015, **348,** 683-686.
2. Guo Z, Manser JS, Wan Y, Kamat PV, Huang L. Spatial and temporal imaging of long-range charge transport in perovskite thin films by ultrafast microscopy. *Nat Commun* 2015, **6,** 7471.
3. Leblebici SY, Leppert L, Li Y, Reyes-Lillo SE, Wickenburg S, Wong E, *et al.* Facet-dependent photovoltaic efficiency variations in single grains of hybrid halide perovskite. *Nat Energy* 2016, **1,** 16093.
4. Stranks SD. Nonradiative Losses in Metal Halide Perovskites. *ACS Energy Letters* 2017, **2,** 1515-1525.
5. Abdi-Jalebi M, Andaji-Garmaroudi Z, Cacovich S, Stavrakas C, Philippe B, Richter JM, *et al.* Maximizing and stabilizing luminescence from halide perovskites with potassium passivation. *Nature* 2018, **555,** 497-501.
6. deQuilettes DW, Koch S, Burke S, Paranji RK, Shropshire AJ, Ziffer ME, *et al.* Photoluminescence Lifetimes Exceeding 8 μs and Quantum Yields Exceeding 30% in Hybrid Perovskite Thin Films by Ligand Passivation. *ACS Energy Lett* 2016, **1,** 438-444.
Page **21** of **23**